\documentclass{llncs}
\usepackage{graphicx} 
\usepackage{comment}
\usepackage{amsmath}
\usepackage{xcolor}
\usepackage{comment}
\usepackage{amsfonts}
\usepackage{siunitx}

\title{A conditioned UNet for Music Source Separation}
\author{Ken O'Hanlon \inst{1,2}
\and Basil Woods \inst{2}
\and Lin Wang \inst{1}
\and Mark Sandler \inst{1}} 
\institute{Centre For Digital Music, Queen Mary University of London
\and AudioStrip Ltd. London}
\date{June 2025}

\begin{document}

\maketitle

\begin{abstract}

In this paper we propose a conditioned UNet for Music Source Separation (MSS).
MSS is generally performed by multi-output neural networks, typically UNets, with each output representing a particular stem from a predefined instrument vocabulary. 
In contrast, conditioned MSS networks accept an audio query related to a stem of interest alongside the signal from which that stem is to be extracted.
Thus, a strict vocabulary is not required and this enables more realistic tasks in MSS.
The potential of conditioned approaches for such tasks has been somewhat hidden due to a lack of suitable data, an issue recently addressed with the MoisesDb dataset. 
A recent method, Banquet, employs this dataset with promising results seen on larger vocabularies.
Banquet uses Bandsplit RNN rather than a UNet and the authors state that UNets should not be suitable for conditioned MSS. 
We counter this argument and propose QSCNet, a novel conditioned UNet for MSS that integrates network conditioning elements in the Sparse Compressed Network for MSS.
We find QSCNet to outperform Banquet by over $1\si{dB}$ SNR on a couple of MSS tasks, while using less than half the number of parameters.

\end{abstract}

\section{Introduction}
\label{sec:intro}

Music Source Separation (MSS) attempts to separate a signal representing a song into several different signals containing the stems of individual instruments present in that song. 
MSS was previously considered a very difficult task with little success beyond specialised cases such as vocal separation \cite{ikala}. 
The introduction of deep learning to MSS has resulted in consistent performance gains
\cite{openunmix} \cite{spleeter} \cite{demucs} \cite{dttnet} \cite{hdemucs} \cite{htdemucs} \cite{bsrnn} \cite{bsrope} \cite{scnet},
particularly in the {\it{four-stem task}} in which the system attempts to separate {\it{vocals}}, {\it{bass}} and {\it{drum}} stems alongside a further catchall {\it{others}} category. 
Separation is often performed by a network with multiple output heads, one per instrument \cite{demucs} \cite{scnet}, or by employing separately trained networks for each instrument \cite{openunmix} \cite{dttnet} \cite{bsrnn}.
Most of these networks employ UNet-based architectures \cite{spleeter} \cite{demucs} \cite{dttnet} \cite{hdemucs} \cite{htdemucs}  \cite{scnet}
that enhance encoder/decoder networks with skip connections between corresponding decoder and encoder modules.
A notable exception are bandsplit networks \cite{bsrnn} \cite{bsrope} in which disjoint spectrogram frequency bands are encoded, and similarly decoded, in a single block. 
Bandsplit RNN (BSRNN) \cite{bsrnn} shares similarities with other recent high-performance networks, such as a dual-path approach in the neck between encoder and decoder  \cite{dttnet}  \cite{scnet} \cite{htdemucs}, and the learning of complex masks that are applied to the input spectrogram to calculate stem approximations \cite{scnet} \cite{dttnet}. 

One reason for the predominance of the four-stem task has been the existence of MusDb \cite{MusDb}, a dataset that is well-formed for the task.
Occasionally this task has been augmented with extra stem categories such as {\it{guitar}} \cite{spleeter}, or {\it{guitar \& piano}}  \cite{htdemucs}.
However, these efforts required private data and still used fixed vocabularies while an ambiguity in some stem categories is also noted \cite{htdemucs}.
Alternatively, conditioned MSS networks \cite{fewshot} \cite{cunet} \cite{upfcond} \cite{septransyn} \cite{lasaft} \cite{leecond} \cite{banquet} possess the potential to avoid such problems through the elimination of hard instrument categories in the network outputs.
In such conditioned networks, the input signal is accompanied by an audio query related to the stem  intended for separation. 
Typically, an embedding representation of the audio query is derived and presented to a separation network using a Feature-wise Linear Modulator (FiLM) \cite{film} layer. 
The FilM layer is placed at some point in a network where it modulates channel activations to emphasise certain features, in this case related to the signal of a particular stem.

Similar to the more general MSS problem, UNet architectures have primarily been used for conditioned MSS. Most of these conditioned UNets to date have employed MusDb \cite{MusDb} as the primary dataset \cite{lasaft} \cite{leecond} \cite{cunet} \cite{fewshot} \cite{zeroass} while the URMP dataset
\cite{URMP} has also been used with similar architectures \cite{upfcond} \cite{septransyn}.
MusDb constrains the possibilities of conditioned MSS as it provides only a small set of four, mostly active, stem categories.
URMP does provide a richer stem vocabulary, but the dataset is small and recorded in a homogenous fashion.
A recent dataset, MoisesDb \cite{MoisesDb} consists of multi-track stems, with  
a stem hierarchy defined with $11$ different stem categories each consisting of subcategories e.g. the guitar  category consists of the sub-categories of \{ {\it clean electric guitar, distorted electric guitar, acoustic  guitar}\}. 
Having such a rich ontology, MoisesDb allows  development of new tasks beyond four-stem separation.
Banquet \cite{banquet}, a conditioned MSS network based upon BSRNN \cite{bsrnn} is one of the first papers to exploit this new resource, with promising results seen on larger vocabularies.
In Banquet, a state of the art music instrument identification network, PASST \cite{passt}, is used to extract query embeddings. 
The queries are presented to a FiLM layer \cite{film} 
placed just before the decoder, similar to \cite{fewshot}. This contrasts to a variety of locations seen in earlier networks; in every decoder block \cite{lasaft}, in every encoder block \cite{cunet} \cite{upfcond}, at every convolutional layer \cite{zeroass} \cite{konguss}.
The adaptation of BSRNN in Banquet is based on a rationale that UNets are not suitable for conditioned MSS as they have problems with information flow \cite{banquet} \cite{ccbanquet}, which are averted with the frequency band split monolithic encoder and decoder layers of BSRNN.

In this paper we consider that the assertion of poor information flow in UNets may not hold, as the skip connections in UNets should help information flow. 
Indeed, most multi-output MSS networks to date are UNets. 
Therefore, we propose the Query-SCNet (QSCNet), a conditioned variant of the Sparse Compressed Network (SCNet) \cite{scnet}, a UNet architecture that performs similar to BSRNN on the four stem task.
We show superior MSS results on MoisesDb for some tasks outlined in the Banquet paper \cite{banquet}, particularly on the $6$ stem problem where a very large improvement of $1.6\si{dB}$ SNR is seen. Meanwhile we observe that QSCNet requires only around $40\%$ of the parameters of Banquet.

\section{Music Source Separation with UNet}
\label{sec:format}

Music source separation  seeks to separate a musical track into constituent
stems that each contain a signal of an instrument or instrument class. 
Consider a musical signal, $\mathbf{y} \in \mathbb{R}^{C \times N} $, with $C$ channels and $N$ samples, and a set of instruments $ \mathcal{I} $ that form the stem vocabulary,
The MSS problem can then be defined as finding the set of sources $\{\mathbf{s}_i \in \mathbb{R}^{C \times N} \} $ such that:
\[
\mathbf{y} \approx \sum_{i \in \mathcal{I} } \mathbf{s_i}.
\]
This is a difficult problem as the membership and cardinality of stems in $\mathcal{I}$ for a given piece may be unknown, and the stems typically outnumber the channels, of which there are $2$ for stereo recordings.
In most MSS efforts to date, the set of instruments employed is typically predefined e.g. $\mathcal{I}^4 = \{bass, vocals, drums, others\}$, and a network $\mathcal{N}$ is applied to a signal leading directly to several corresponding outputs
\begin{equation}
\label{MSN}
\mathcal{N}(\mathbf{y}) \longrightarrow \{\mathbf{  s_i\}_{i\in{\mathcal{I}}}}
\end{equation}
although in some cases one network is trained for each instrument
\begin{equation}
\label{1SN}
\mathcal{N}_i(\mathbf{y}) \longrightarrow \mathbf{s_i}.
\end{equation}

The UNet \cite{unet} is a commonly used architecture in MSS problems, usually applied to the complex spectrogram: $\mathbf{X} \in \mathbb{C}^{F \times T} = \mathrm{STFT}(\mathbf{y})$. 
UNets can be considered to comprise several common subnetworks, 
the encoder $\mathcal{N}^{Enc}$, decoder $\mathcal{N}^{Dec}$, and the neck connecting these two subnetworks $\mathcal{N}^{Neck}$. 
While the original UNet was fully convolutional, variants employed in MSS typically cast the $\mathcal{N}^{Neck}$ as an RNN-based\cite{hdemucs} \cite{scnet}, or transformer-based \cite{htdemucs} module. 

Similar to standard encoder / decoder architectures, the encoder in the UNet consists of several, $L$, sequential modules that reduce the feature dimension
$\mathcal{N}^{Enc} = (\mathcal{N}^{e_l} )_{l \in \mathcal{L}}$ where $\mathcal{L} = \{1, .., L\}$
while the decoder similarly consists of several modules $\mathcal{N}^{Dec} = (\mathcal{N}^{d_l} )_{l \in \mathcal{L}}$  that increase the feature dimension.
In time-frequency domain UNet MSS the encoder and decoder modules typically only alter the dimension in the frequency direction while the size of the temporal dimension remains constant through the network, although there are exceptions to this \cite{dttnet}.
A defining feature of UNets is that the encoder and decoder modules with similar representation dimensions are joined by skip connections.
Given that $\mathbf{e_l} = \mathcal{N}^{e_l} (\mathbf{e_{l-1}})$ describes the output of the $l$th encoding module, the operation at the $(L-l)$th decoding module is then described as 
$\mathbf{d_l} = \mathcal{N}^{d_l} (\mathbf{d_{l-1}, e_{L-l}})$.

The specific case of a mask based MSS UNet can be considered as the sequence of operations
\begin{equation}
\label{uneteq}
\begin{aligned}
\mathbf{E} &= \mathcal{N}^{Enc}(\mathbf{X})  \\ 
\mathbf{N} &= \mathcal{N}^{Neck}(\mathbf{E}) \\
\{\mathbf{M_i} \} &= \mathcal{N}^{Dec} (\mathbf{N}, \{ \mathbf{e_l} \}_{l \in \mathcal{L}})  \\
\{ \mathbf{s_i} \} &= \{ \mathrm{ISTFT} (\mathbf{M_i} \otimes \mathbf{X} ) \}_{i \in \mathcal{I}}
\end{aligned}
\end{equation}
where $\mathbf{M}_i$ denotes the complex mask related to the $i$th stem in $\mathcal{I}$ and $\mathbf{E} = \mathbf{e_L}$.

\subsection{Conditioned Music Source Separation}

An alternative approach to MSS does not require a fixed stem vocabulary \eqref{MSN} or an ensemble of stem-wise networks \eqref{1SN}. 
Conditioned approaches to MSS employ one network for all possible instruments and supply either a representation of category \cite{premcond} or an audio query, $\mathcal{Q}$, as input alongside the input signal:
\begin{equation}
\label{conet}
\mathcal{N}(\mathbf{y}, \mathcal{Q}_i) \longrightarrow \mathbf{s_i}.
\end{equation}
Such an approach may also be referred to as query-based source separation.
While $\mathcal{Q}$ can be a one-hot vector with each dimension representing the activity of a given stem \cite{premcond}, it is more flexible to supply $\mathcal{Q}$ as a query embedding that represents a point, or area, in an audio feature space. 
The query embedding is usually output from a separate neural network such as one trained for musical instrument recognition, 
and typically derived from the activations in the penultimate layer of the network.

A Feature-wise Linear Modulation (FiLM) \cite{film} module is employed to condition the network activations $\mathbf{P} \in \mathbf{R}^{C \times F' \times T'} $ at some point in the network based upon a query embedding, $\mathcal{Q} \in \mathbb{R}^{q}$
\begin{equation}
\label{filmeq}
\mathbf{P} \longleftarrow \mathcal{N}^{FilM}( \mathcal{Q},  \mathbf{P} ).
\end{equation}
The FiLM module, $\mathcal{N}^{FiLM}$ learns an affine transformation function with parameters $\mathbf{o_{\gamma}}, \mathbf{o_{\beta}}$ that are typically small neural modules that respond to the vector query input $\mathcal{Q}$:
\begin{equation}
\label{filmops}
\gamma = \mathbf{o_{\gamma}} (\mathcal{Q}); \qquad \quad \beta = \mathbf{o_{\beta}} (\mathcal{Q})
\end{equation}
with the resultant vectors  $\gamma, \beta \in \mathbb{R}^C$ applied to the input representation
\begin{equation}
\label{filmeff}
\mathbf{P}_{c,f,t} \longleftarrow \gamma_c \mathbf{P}_{c, f, t} + \beta_c.
\end{equation}

As well as the query-based modulation, conditioned MSS differs from the multi-stem based approach \eqref{uneteq} at the mask formation and output stage. 
At this point, the set of masks has only one member, $\mathbf{M}$, resulting in only one output signal, $\mathbf{s}$.
Nevertheless, multi-stem separation can still be performed in one batch by inputting several queries with one signal and performing post-FiLM processing as a batch. 
In this way, the computational load at inference time may be reduced as the signal encoding is performed only once.

\section{Proposed Approach}

We propose a conditioned UNet for MSS called Query-SCNet (QSCNet).
QSCNet is an adaptation of the Sparse Compressed Network \cite{scnet} (SCNet) which we embellish with conditioning capabilities. 
Our rationale for adopting SCNet includes some perceived similarities to BSRNN, from which Banquet is adopted. 
BSRNN and SCNet are seen to perform similarly in the four-stem MSS task \cite{scnet}, and both possess features found in modern MSS networks such as dual-path RNN and complex mask learning.
A further consideration in the selection of SCNet is its relatively small size compared to other state-of-the-art MSS networks \cite{scnet}.

\subsection{SCNet}

The authors of the original paper \cite{scnet} do not describe SCNet as a UNet.
Rather, they opt to convey an additional fusion layer that accepts similar inputs as a UNet decoder module, $(\mathbf{e_l}, \mathbf{d_{L-l}})$ and outputs to the decoder.
However, we state here that SCNet can be formulated as a UNet variant \eqref{uneteq} simply by aggregating the fusion and decoder layers as they are defined in the paper \cite{scnet}.
In this light, the main point of difference of the SCNet from other UNets is the novel proposed banded downsampling and upsampling modules employed in the encoder and decoder, and a novel dual-path RNN.
We briefly outline these here as QSCNet naturally inherits these features.

In SCNet the stereo complex spectrogram is first gathered into a tensor with $4$ channels.
At each time-frequency point in this tensor the $4$ channels represent the corresponding complex STFT's real and imaginary   coefficients across the two stereo channels \cite{scnet}.
At each subsequent block of the encoder the inputs are subjected to a coarse banding, with fixed ratios, across the frequency dimension. 
This results in $3$ separate banded tensors, to which different downsampling ratio and convolutional processing strategies are applied, before they are regathered at the output of the encoder block.
Although fixed ratios are used in the banding, it is notable that this does not imply consistent frequency band splitting across the full decoder as the banding is only applied to a dimension with a linear frequency scale in the first downsampling layer.
A corresponding recombination strategy is effected in the decoder.

The neck of SCNet consists of $6$ dual-path bidirectional LSTMs that are processed in alternating fashion. Generally in MSS \cite{dttnet} \cite{bsrnn} a dual-path RNN refers to first processing across the temporal dimension using a RNN before similar processing across the frequency dimension. 
A novel variation is present in the neck of SCNet, where alternating dual-path RNNs operate on different domains \cite{scnet}. 
The first and subsequent odd numbered dual-path RNNs operate directly on the latent feature, while even numbered dual-path RNNs operate on the Fourier representation of the latent feature.
Between the odd- and even-numbered dual-path RNNs, the real FFT is applied to the feature map, and processing is performed on the latent feature in its frequency domain. After this processing is performed, between the even and odd numbered dual-path RNNs, the inverse RFFT is applied to bring the Fourier-based feature map back into the original latent feature domain \cite{scnet}.

\subsection{Conditioning}

Embedding-based queries are used for QSCNet. 
Similar to \cite{banquet}, we employ the PASST network \cite{passt} in order to extract embeddings, thereby maintaining similarities  between QSCNet and Banquet.
Specifically, we used the variant of PASST that was trained on the OpenMic \cite{openmic} dataset for the instrument recognition task and is freely available from the authors \cite{passt}. 
This version of the network only has $20$ outputs representing different instruments, or instrument categories and again similar to \cite{banquet} we remove the final layer and employ the outputs as an embedding $\mathcal{Q} \in \mathbb{R}^{768}$ given an audio clip of up to $10\si{s}$.

\begin{figure}
\includegraphics[width=\textwidth]{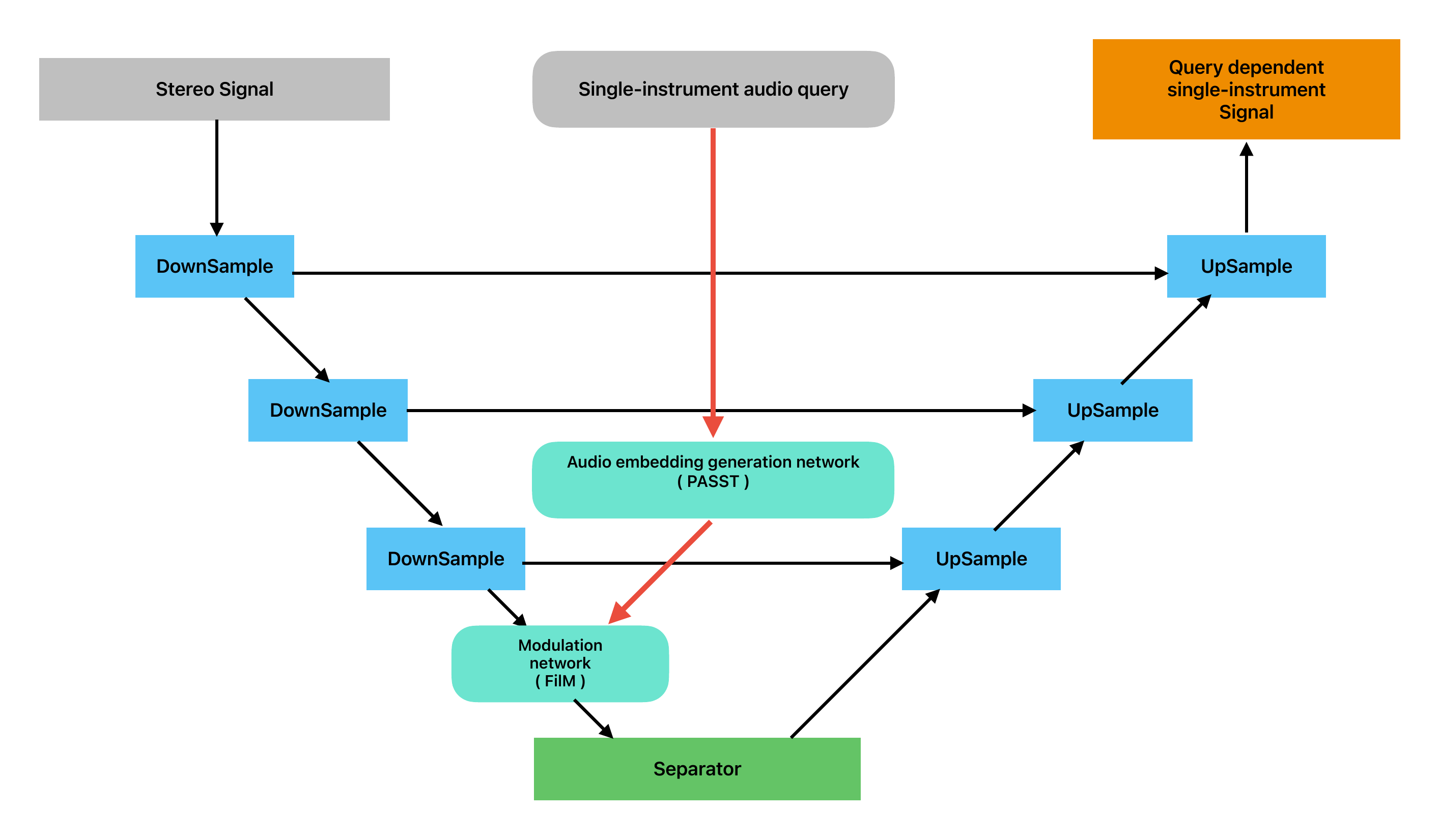}
\caption{A schematic diagram of QSCNet, or a generic UNet, with L=3, a PASST embedding generator network, and a FiLM modulator network integrated at end of encoder. } \label{fig1}
\end{figure}

We employed just one FiLM module, which we located at the end of the encoder and directly before the dual-path RNN as can be seen in Fig. 1. Otherwise put, in \eqref{filmeq} we set $\mathbf{P} = \mathbf{E}$, where $\mathbf{E}$ is the encoder output \eqref{uneteq}.
Although we are aware that this is an unused location for FiLM in conditioned MSS, 
we consider that this position should be optimal for conditioning as it allows the instrument context to be defined before the sequential long-term processing.
We recall that several works have used different conditioning locations, and have used several FilM modules at different locations.
Perhaps of most interest here, in Banquet the conditioner is placed at the end of $\mathcal{N}^{Neck}$, just before start of the decoder.

The FiLM module employed here uses similar networks for the parameters $\mathbf{o}_\gamma, \mathbf{o}_\beta$ \eqref{filmops} of the affine transformation \eqref{filmeff}. 
These networks consist of multi-layer perceptrons with two fully connected layers; the first layer has $q = 768$ inputs and $c = 128$ outputs, followed by an ELU activation \cite{ELU}; the second layer has $c=128$ inputs and outputs $\gamma$ or $\beta$, respectively \eqref{filmops}.

\section{Experiments}
\label{sec:majhead}

We ran experiments to test the efficacy of the proposed QSCNet.
We trained QSCNet on a $6$ stem vocabulary 
$\mathcal{I}^6 = \{vocals, bass, drums, guitar, piano, others\} $ similar to that employed in\cite{htdemucs}.  
We also propose a six-stem variant of SCNet (SCNet6) which is also trained on the same vocabulary. 
$\mathcal{I}^6$ contains the same instrument stems as the Q:VDBGP vocabulary employed in \cite{banquet}, which allows a direct comparison with Banquet. However, $\mathcal{I}^6$ also contains the 
$others$ category, which allows QSCNet and SCNet6 to be compared more directly.

As the conditioned network model can take any input query and return a corresponding output \eqref{conet} there is a flexibility where, unlike in multi-stem networks, a model trained with one vocabulary can be tested on another. 
In this light, we also test on an extended variant of $\mathcal{I}^6$ that includes some finer stems e.g. {\it male vox} and {\it female vox} rather than just the coarser {\it vocals}.

\subsection{Data}

We used MoisesDb \cite{MoisesDb}, which consists of $240$ songs with multi-track audio with the training/validation/test splits proposed in \cite{banquet} with $144/48/48$ songs, respectively.
MoisesDb uses a hierarchial structure with $11$ separate stem categories, each of which may contain finer stem subcategories that number $30$ in all.
However, unlike $4$-stem datasets like MusDb, the data distribution is skewed.  \cite{MoisesDb}. 
Some stems such as bass, drums and vocals from $\mathcal{I}^4$ are present in almost every song, while some instrument categories do not appear in all datasplits \cite{banquet}.

We first constructed the training dataset employing the vocabulary $\mathcal{I}^6$. 
For each stem in this category, all its finer subcategory tracks were added to form a single stem track.
The $others$ stems were then formed from all remaining tracks not assigned to one of the instrument categories. 
The validation and test sets were formed in a similar fashion.

We employed a random data sampling strategy at training time in which training clips were generated by mixing randomly selected clips for the different stems.
Such cacophonous mixtures of unrelated stems of music have been shown to be superior for training neural networks for MSS \cite{cacophony}. 
Audio clips of $10\si{s}$ were used to train QSCNet and the $6$ stem SCNet.
First, a set of candidate clips was assembled for each stem by selecting $10\si{s}$ audio segments spaced $1\si{s}$ apart in each audio track. 
Each candidate stem clip is accepted into the final pool of clips after an inspection for a simple silence detection that rejects segments in which more than $50\%$ of samples are zero.

At train time, an audio clip is randomly selected for each stem category.
Some augmentation is applied to the stem clip data. This includes channel flipping in which the stereo channels for an instrument are swapped, and sign flipping where the sign of all elements of a signal are swapped. Both of these flipping augmentations are activated randomly with even chance.
A gain augmentation is also applied in which each individual instrument stem segment 
is subjected to the application of a gain randomly selected between in the range $(0.25, 1.25)$. This is similar to the augmentation setup used in the Demucs networks \cite{htdemucs} and in SCNet \cite{scnet}. 
The training mixture is then formed by mixing the various augmented stem clips together.

For the conditioned approach, a pool of queries for each instrument was also  formed. Here, this is generated simply by further filtering of the audio clip pools used in training above.  Specifically, only audio clips in which less than $20\%$ of samples are zero are employed as queries.
In the conditioned training, one query instrument is selected from $\mathcal{I}^6$ with an even chance of each instrument being selected. A random query from the pool is then selected, and input to the network alongside the augmented audio mixture.
Similarly, at validation and test time a random query is selected for each instrument from its query pools in the validation or test set, respectively. 

A test set for $\mathcal{I}^6$ was formed using the subcategory gathering strategy for stems, as above.
An extra test set was formed from the same data using an extended vocabulary, referred to as $\mathcal{I}^{6E}$, that substitutes the {\it vocals, guitar \& piano} categories with finer stems.
Specifically the vocals are split into {\it male \& female vocals}, the guitar is split into {\it clean electric guitar, distorted electric guitar \& acoustic guitar }, while the piano is represented by the {\it grand piano \& electric piano} subcategories. This results in a $10$ stem category vocabulary, if the $others$ category is included.

\subsection{Parameters}

Spectrograms used for network inputs were produced from stereo audio files sampled at $44.1\si{kHz}$ using a window size of $4096$ with $75\%$ overlap.
The root mean square energy (RMSE) was used as a cost function, as for SCNet \cite{scnet}. 
Experiments were run on a single $A100$ chip with $80\si{Gb}$ of memory using a batch size of $8$. 
Some initial experiments were run with batch sizes of $4$ and $16$ We found that training was sometimes poorer with a batch size of 4, while training was slower, requiring more epochs, when the batch size was $16$.
In each batch $32000$ samples were taken resulting in $4000$ mini-batches.
The Adam \cite{adam} optimizer was employed for training with the learning rate set to $3\times10^{-4}$. 
Each model was trained for $300$ epochs, with validation performed after each epoch.
Similar to \cite{htdemucs}\cite{scnet} exponential moving averages were maintained after each epoch. The model, or averaged model, that performed best on the validation set was kept as the final trained model.

\subsection{Metrics}

On each track, the signal-to-noise ratio for the $i$th instrument was calculated
\[
SNR_i = 10 \times \log_{10} {\frac{\|\mathbf{y_i}\|_F^2}{\|\mathbf{y_i - s_i} \|_F^2}}
\]
where $\mathbf{y_i}$ and $\mathbf{s_i}$ are the known and approximated stem signals respectively.
For each instrument the median SNR across all tracks of the same instrument is recorded. 
This is similar to the metric used in \cite{banquet} which enables a direct comparison of results. 

\begin{table*}[t]
  \centering
  \begin{tabular}{r| c c c c c |c| c}
    Alg & Bass & Vocals & Drums & Guitar & Piano & Avg5 & Others  \\
    \hline
    Banquet & 11.0 & 8.0 & 9.5 & 3.3 & 2.5 & 6.9 & - \\
    QSCNet & 11.9 & 9.8 & 11.7 & 5.7 & 3.4 & 8.5 & 1.3 \\
    \hline
    HTDemucs & 10.9 & 8.9 & 11.6 & 2.4 & 1.7 & 7.1 & -  \\
    SCNet6 & 12.8 & 10.5 & 12.4 & 6.3 & 4.0 & 9.2 & 2.8\\
    SCNet6(L) & 13.5 & 12.2 & 13.4 & 7.0 & 4.6 & 10.1 & 3.4 \\
    \hline
    
    \hline
   
\end{tabular}
  \caption{Results comparing 6 stem approaches HTDemucs6, SCNet6 \& SCNet6(L), and two conditioned approaches, Banquet \cite{banquet} and QSCNet for MSS on the MoisesDb test set. Instrumentwise results given in median SNR. }
  \label{tab:1}
\end{table*}

\subsection{Results}

The results on $\mathcal{I}^6$ are shown in Table $1$, where QSCNet is compared with the proposed SCNet6 and its larger variant, SCNet6(L), which has the same architecture as SCNet6 but with double the number of channels in each layer. 
Here they are also compared to state-of-the-art results given in \cite{banquet} for the six stem HT-Demucs \cite{htdemucs} and the conditioned Bandsplit network, Banquet \cite{banquet} both of which are evaluated for $5$ stems only, 
as the {\it others} category was not considered \cite{banquet}.
For each algorithm an average score over the $5$ instrument stems (Avg5) is also recorded in order to afford a simple summary comparison.

In terms of the multi-output networks SCNet6 is seen to outperform HT-Demucs by a large margin of $2.1\si{dB}$, with improvements for all instruments. Some of these improvements are very large e.g. for the guitar there is an increase of $3.9\si{dB}$. This is perhaps more impressive when it is considered that SCNet6 is trained only on $144$ songs of MoisesDb while HT-Demucs was trained on a private dataset of $800$ songs \cite{htdemucs}. 
Further improvements are seen with SCNet6(L) with the Avg5 metric $3\si{dB}$ above the HTDemucs. 
Large improvements using SCNet6(L) relative to the standard SCNet6 are
seen for the vocals and drums categories.

Considering the conditioned networks, the proposed QSCNet is seen to be superior to Banquet by a large margin of $1.6\si{dB}$ on the Avg5 metric. 
Improvements are seen across all instruments when using QSCNet, notably on drums and guitar which both improve over Banquet by more than $2\si{dB}$.
The QSCNet is also seen to improve on the HTDemucs $6$ stem, and reach a performance level around $0.7\si{dB}$ lower than that of SCNet6.
It is notable that QSCNet uses many less parameters that SCNet6, as many parameters are employed in the formation of the individual masks.
We observe that QSCNet uses $10.2$M parameters; while SCNet4 and SCNet6 employ $20.4$M and $26.6$M parameters respectively. QSCNet is also significantly smaller than Banquet which uses $24.9$M parameters.

\begin{table*}[b]
  \centering
  \begin{tabular}{r c c c c c c c c c c c}
    \cline{4-10}
    \multicolumn{1}{c}{} & \multicolumn{1}{c}{}  & \multicolumn{1}{c}{}  & \multicolumn{2}{|c|}{Vocals} & \multicolumn{3}{c|}{Guitar} & \multicolumn{2}{c|}{Piano} & \multicolumn{1}{c}{}\\
    \multicolumn{1}{c}{} & Drums & Bass & \multicolumn{1}{|c}{Male} & \multicolumn{1}{c|}{Female} &  \multicolumn{1}{|c}{Acoust.} &  \multicolumn{1}{c}{Clean}  & \multicolumn{1}{c|}{Dist.} & \multicolumn{1}{c}{Grand} & \multicolumn{1}{c|}{Elec.} & Avg9 & Others \\
    \hline
    \multicolumn{1}{r|}{Banquet} & 10.1 & 10.7 & 7.9 & 10.1 & 0.9 & 1.7 & 2.8 & 2.8  & 0.5 & 5.3 & - \\
    \hline
    \multicolumn{1}{r|}{QSCNet} & 11.8 & 11.6 & 8.5 & 11.8 & 1.3 & 3.6 & 4.0 & 3.2 & 0.7 & 6.3 & 1.1\\
    \hline   
  \end{tabular}
  \caption{Results comparing Banquet\cite{banquet} and the proposed QSCNet for MSS on the MoisesDb test set for the extended $\mathcal{I}^{6E}$ vocabulary. Instrumentwise results given in median SNR. }
  \label{tab:1}
\end{table*}

Results on the extended $\mathcal{I}^6$ are shown in Table 2 where QSCNet is seen again to improve over Banquet, this time by an average of $1\si{dB}$, and without relative degradation for any instrument. 
In this case it is notable that Banquet has been trained on the fine stem vocabulary  $\mathcal{I}^{6E}$ that both algorithms are tested on, while QSCNet was trained on the coarse stem version, $\mathcal{I}^6$.

\section{Conclusions}

We proposed the QSCNet, a conditioned variant of SCNet\cite{scnet}.
Experimental results show that this method improves significantly over the Banquet model, with state of the art results seen for conditioned MSS on the Q:VBDGP problem on both coarse and fine stems.
We think that this validates the adoption of SCNet, and the selected location of the FiLM element in QSCNet.
In producing these results using only one FiLM element and a network that is substantially smaller than Banquet, we would argue that we have solidly refuted any negative assertions about the suitability of UNets for conditioned MSS presented in \cite{banquet}. 

We also considered the $6$ stem problem by training a SCNet(6) with MoisesDb on a six stem problem.   State of the art results were recorded with SCNet6 and SCNet6(L) strongly outperforming the HTDemucs6 although trained on a much smaller dataset. Interestingly the performance of QSCNet in terms of $6$ stems was $ \SI{0.7}{dB}$ less than the SCNet6.
This is interesting as the SCNet6 requires $250\%$ the number of parameters of QSCNet although it uses 
similar size network elements. We consider that the performance gap between multi-stem and conditioned methods might further reduce if using a similar number of parameters. 

Although performance improvements seen here are substantial, we do consider they may not be solely due to the choice of base network that is being conditioned. 
Future work will perform some ablation of the proposed approach, with
some preliminary results not shown here suggesting that some differing network and training parameters may contribute somewhat to the performance gap.
Furthermore we shall look at training on larger vocabularies such as found in \cite{banquet}, and considering performance for conditioned and multi-output networks for a similar number of parameters.

We believe that this is a particularly timely contribution.
Conditioned MSS provides capabilities that were previously unavailable in MSS.
These capabilities have rarely been fully exposed as most research in conditioned MSS was applied to small fixed vocabulary data, such as MusDb, to which the approach is not really suited.
Now, with MoisesDb there is now a research dataset available that should encourage more research in conditioned MSS.
Indeed, we think conditioned MSS should, become the most focussed area of research in MSS due to its potential capabilities.
We think that the work presented here is a positive step in this direction.

\bibliographystyle{splncs04}
\bibliography{refs}

\end{document}